\documentclass[preprint,authoryear,12pt]{elsarticle}

\usepackage{graphicx}
\usepackage{amssymb}

\journal{Icarus}

\begin{document}

\begin{frontmatter}

\title{Saturn Ring Seismology: Looking Beyond First Order Resonances} 

%% use optional labels to link authors explicitly to addresses:
%% \author[label1,label2]{<author name>}
%% \address[label1]{<address>}
%% \address[label2]{<address>}

\author{Mark S.\ Marley}

\address{Mail Stop: 245-3, NASA Ames Research Center, P.O. Box 1, Moffett Field, CA 94035-0001}

\begin{abstract}
%% Text of abstract
Some wave features found in the C-ring of Saturn appear to be excited by resonances with normal mode oscillations of the planet. The  waves are found at locations in the rings where the ratio of orbital  to oscillation frequencies is given by $m : m+1$ where $m$ is a small integer.  I suggest here that it is plausible that ring waves may also be launched at second order resonances where the frequency ratio would be  $m : m+2$. Indeed otherwise unassociated wave features are found at such locations in the C-ring. If confirmed the association of planetary modes with additional C-ring wave features would measure additional oscillation frequencies of Saturn and improve the utility of the waves for constraining the internal structure of the planet. Second-order resonances in general do not lie near first order ring resonance locations and thus are not the explanation for the  
apparent frequency splitting of modes.
\end{abstract}

\begin{keyword}
Saturn\sep rings\sep seismology
%% keywords here, in the form: keyword \sep keyword

%% MSC codes here, in the form: \MSC code \sep code
%% or \MSC[2008] code \sep code (2000 is the default)

\end{keyword}

\end{frontmatter}

% \linenumbers

%% main text
\section{Introduction}
\citet{M90} and \citet{MP93} proposed that certain wave features in Saturn's C-ring discovered by \citet{1991Icar...93...25R}, as well as the Maxwell gap, were created by resonant interaction with internal oscillation modes of Saturn. Since the oscillation modes perturb the internal density profile and consequently the external gravitational field of the planet, they have the potential to excite vertical and horizontal excursions in ring particle orbits analogously to the forcing applied by a  satellite of the planet.  The precise radial location of the ring features depends on the planetary oscillation mode frequency and thus the rings can serve as a seismometer, recording the frequencies of Saturnian oscillation modes.

While Marley \& Porco (1993) argued that the Rosen waves could be associated with Saturnian oscillation modes\footnote{\citet{1987BAAS...19..889M} originally suggested that Saturn f-modes might launch C-ring waves prior to the announcement of the discovery of unassociated waves by \citet{1988BAAS...20..853R}. }, their detailed predictions for the characteristics of the waves expected to be excited in the rings could not be tested by the Voyager data available at the time.  In particular the 
Voyager Radio Science dataset employed by \citet{1991Icar...93...25R} (hereafter R91) was not adequate to precisely ascertain the azimuthal wavenumber, $m$, of the ring features beyond the crude constraint that $m$ was generally small, in the range of 2 to 5 or so. Furthermore while the direction of propagation of the waves was apparent, the Voyager data were not sufficient to ascertain whether  the R91 waves were density or bending waves.

Such details were important since waves launched by interactions with Saturnian internal modes would behave differently than waves launched by external satellites.  Density waves launched at inner Lindblad resonances (ILR) with external satellites propagate outwards--away from Saturn--while bending waves launched at inner vertical resonances (IVR) propagate inwards. If any of the wave features identified in R91 were indeed produced by resonances with the planet, then the direction of propagation would be the opposite of the satellite case.   Bending waves should propagate outwards and density wave inwards. Since neither the  wave type (bending or density) nor their azimuthal wavenumber were constrained by R91, the ring waves could thus not be definitively associated with oscillation modes of the planet.  The situation did not change for 20 years.

Recently, however, \citet{HN13} used stellar occultation data obtained by the Cassini Visual and Infrared Mapping Spectrometer to place new constraints on the C ring waves first identified by R91. They focused on six waves, four of which had first been identified by R91, and concluded that they were density--not bending--waves and that their azimuthal wavenumbers were consistent with the predictions of \citet{MP93} (hereafter MP93). \citet{HN13} thus concluded that a seismological origin for these waves was likely, thereby confirming the MP93 hypothesis.  One surprise, however, was that there were multiple waves with the same azimuthal wavenumber $m$ at different locations in the C-ring, whereas only one wave would be expected.  The usual mode splittings arising from rotation and planetary oblateness has already been accounted for in the mode frequency calculation of \citet{M91}. Thus, some other property of Saturn must permit multiple oscillation modes, each with the same azimuthal wave number, but with slightly different frequencies. 

\citet{Ful13} examined the possible role of mixing between the global f-mode oscillations of Saturn and modes that might be trapped in Saturn's core.  They found that such mixing could in principle create multiple modes with similar frequencies near a single f-mode and explain the apparent mode splitting, but that a substantial `tuning' of the Saturn envelope and core model would be required to produce the observed ring features and they deemed this unlikely.  The origin of this ``fine'' mode splitting thus remains unknown.                                                                         

In addition to the waves studied by \citet{HN13}, \citet{2011Icar..216..292B} also searched Cassini UVIS data for wave features.  By summing multiple stellar occultation profiles they could find waves with lower amplitudes than previous studies.  They tabulated a total of 40 C-ring wave features, of which only five were associated with possible satellite resonances.  With six of the waves apparently of seismic origin a remainder of about 30 could not be explained. The ring wave features tabulated by \citet{2011Icar..216..292B} that are not associated with known satellite resonances are plotted in Figures 1 and 2. Figure 1 shows those features which appear to propagate inwards, towards Saturn while Figure 2 shows those that propagate in the opposite direction. 

Given the discoveries by \citet{HN13}  and  \citet{2011Icar..216..292B}  it seems appropriate to revisit the seismological connection of Saturnian oscillation modes with ring wave features with the aim of testing if planet-ring system resonances beyond those considered by MP93 could account for either the observed apparent fine splitting of modes or any of the waves tabulated by \citet{2011Icar..216..292B}.  To address these issues, here I briefly review the theory presented in MP93 and point out that second order resonances can produce many more additional resonant features in the C-ring than were tabulated in MP93.

\section{Basics of Ring Seismology}
The theory of planetary mode ring seismology is laid out in detail in MP93.  This section provides a very brief summary of the planetary oscillation modes, ring wave excitation mathematics, and explains the calculation of
locations of waves launched by a given planetary oscillation mode.  

\subsection{Which Modes?}

Several different types of planetary oscillation modes, including p-, g-, and f-modes, which differ in the nature of their restoring force (see \citet{Unno79}), could in principle launch waves in the rings. A single planetary oscillation mode as observed from inertial space has frequency commonly denoted by $\sigma_{\ell m n}$. The three integers, $m$, $\ell$, and $n$  uniquely identify a specific mode\footnote{Strictly speaking the index $\ell$ defines a spherical harmonic while the planetary modes on a rotating planet consist of a superposition of several spherical harmonics (see discussion in Section  4.2). For simplicity we here equate a single f-mode to a single spherical harmonic.}. The index $\ell$ enumerates the number of circles bounding regions of positive or negative perturbation at the surface; $\ell - |m|$ is equal to the number of boundaries in latitude. By the convention used here modes propagating in the same (opposite) direction of rotation of the planet have positive (negative) values of $m$.  The number of radial nodes from the surface to the center of the planet is denoted by $n$.  

As described in MP93 the modes most likely to produce features in the C-ring are f-mode oscillations which are defined as those with no radial nodes, or $n=0$.  This is because the perturbation to the external gravitational field will be greatest when the density perturbation inside the planet is in phase from surface to core. When $n>0$ the radial density perturbation is no longer in phase from core to atmosphere, resulting in a reduced perturbation to the external gravitational field of the planet.  In addition, the frequencies of the $n>0$ modes are too high to resonate with ring particle orbits. As $\ell$ increases the f-modes are trapped progressively closer to the surface where the background density is smaller.  A given mode amplitude at the surface thus produces progressively smaller perturbations to the external gravitational field as $\ell$ is increased.  Ring seismology thus favors modes of low $\ell$.

A complete calculation of the frequency of a planetary oscillation mode requires computing $\sigma^0_{\ell m n}$, the mode frequency in the frame of the planet, as well as corrections for rotation to give the frequency  $\sigma_{\ell m n}$ as observed in inertial space.  The correction must account both for the change out of Saturn's rotating frame (rotation frequency $\Omega_{\rm Sat}$) and the influence of rotation on the modes themselves. For slow rotation the former correction is simply $m\Omega_{\rm Sat}$ and the latter is typically given as $C_{\ell n}$ such that the total correction is written as
$$\sigma_{\ell m n} = \sigma^0_{\ell m n} + m (1-C_{\ell n})\Omega_{\rm Sat }. \eqno(1)$$
Thus solely because of the frame shift a mode rotating prograde ($m>0$) will appear to an observer at rest to have a higher frequency than to an observer in the rotating frame.
Both Jupiter and Saturn rotate so rapidly, however, that the planet is oblate and a more complex rotation correction--to at least second order in $\Omega_{\rm Sat}$--must be applied \citep{VZ81}. However Eq. (1) is conceptually adequate for discussion purposes. 

Note that  in principle a mode propagating in the opposite direction of Saturn's rotation ($m<0$) could be carried prograde if $\sigma^0_{\ell m n}$ was slow enough.
In practice however $\sigma^0_{\ell m n} > \Omega_{\rm Sat}$ and those modes propagating opposite to Saturn's direction of rotation are never carried prograde by rotation and thus cannot resonantly interact with prograde orbiting ring particles. Because of the periodicity of the oscillation perturbation in azimuth, the pattern frequency of the perturbation $\Omega_{\rm pat}$ as seen in inertial space is equal to $\sigma_{\ell m n} / m$. It can sometimes be helpful to think of a mode's ``pattern period'', $P^{\ell m n}_{\rm pat}$, and these are tabulated as well in Table I.  

Launching a density wave in the rings requires that there be azimuthal variations in the external gravitational field around the planet as sensed at the equator.  By the properties of spherical harmonics this means that only those planetary oscillation modes with either $m=\ell$ or $\ell-|m|$ equal to an even integer can excite density waves in the rings.  Other modes will not produce a horizontally varying gravitational potential at the equator. In contrast outer vertical resonances require a vertical gradient in gravitational potential at the equator which in turn implies that $\ell - |m|$ is an odd integer.  

In summary if the oscillation modes of Saturn are to launch waves in the rings, the most plausible modes are low $\ell$ f-modes ($n=0$) with $m>0$.  But where in the rings will the resonances be found?

\subsection{Resonance Locations}

Given these practical constraints, MP93 derived the expected location for eccentric resonances between planet modes and ring particle orbits. They found that the orbital frequencies $\Omega$ associated with  Lindblad resonances were given by
$${\Omega_{\rm pat}\over \Omega} = 1 \mp {q \over |m|} {\kappa \over \Omega}. \eqno(2)$$ 
Here $\Omega$ and $\kappa$ are the orbital angular and epicyclic  frequencies of a ring particle at a given location in the rings and $q$ is a positive integer. Outer (inner) resonances take the lower (upper) sign.  In the case of resonances with external moons, the well known  Lindblad Resonances with $q=1$ are the strongest resonances and MP93 assumed $q=1$ for their tabulation of C-ring wave features. 

In the case of vertical resonances the commensurability is with ring particle vertical excursions of frequency $\mu$,
$${\Omega_{\rm pat}\over \Omega} = 1 \mp {b \over |m|} {\mu \over \Omega}. \eqno(3)$$ 
Note that the radial and vertical epicyclic frequencies, $\kappa$ and $\mu$, are similar to $\Omega$ but the three frequencies not precisely equal for an oblate planet. Again outer resonances arise from the lower sign and $b$ is a positive integer.  MP93 again only considered the case of $b=1$.

For heuristic purposes it is useful to ignore the oblateness of Saturn and set $\Omega = \kappa = \mu$ and then the resonant locations are found at a radial distance $r$ in the rings where
$$\Omega_{\rm OVR}(r) = {\Omega_{\rm pat} \over { 1 \mp {b\over m}}} ~ {\rm and}~ \Omega_{\rm OLR}(r) = {\Omega_{\rm pat} \over { 1 \mp {q\over m}}}. \eqno(4)$$ 
Here OLR and OVR refer to outer Lindblad and outer vertical resonances respecitvely.
For first order outer (`out') resonances ($b=q=1$) we then have $\Omega_{\rm out} : \Omega_{\rm pat}= 1:2$, $2:3$, $3:4$, and $4:5$ for $m=1$ through 4.
MP93 found that for such frequency ratios resonances would be expected to be found in  the C ring and D ring for those modes with $m=\ell -1$ (OVR) and $ m=\ell$ (OLR). Their Table I summarized the most likely candidates. 

If the requirement that $b$ and $q=1$ is relaxed then a new set of frequency ratios appear. For $b$ or $q=2$ such commensurabilities as $\Omega_{\rm out} : \Omega_{\rm pat}= 1:3$, $2:4$, $3:5$, and $4:6$ for $m=1$ through 4 are allowed. Such ratios enable higher frequency modes, which would not produce a first order resonance in the rings, to now potentially launch a wave at a second order resonance. Modes which have a first order resonance in the C-ring would have a second order resonance farther out in the rings.

\section{Higher Order Resonances}

As discussed in Section 2.2, MP93 focused on the locations of first order resonances to present the ring seismology concept. In their Section III.A MP93 briefly considered the possibility of resonances with $b$ or $q>1$  and noted that such resonances would occur ``throughout the C-ring''. However since such resonances would presumably be weak and more difficult to identify, MP93 dismissed them without further discussion. Given that no first order resonances had yet been found this was a conservative choice. However, now that \citet{HN13} have demonstrated that there are indeed likely C-ring features excited by planetary modes, it is worth considering whether second order resonances could also excite ring wave features and where such waves might lie.  In this section we consider the locations of higher order resonances with the same f-modes considered by MP93. 

\subsection{Locations}
In the case of higher order resonances either $b$ or $q$ will be an integer greater than one.  Since the aim here is to demonstrate the applicability of such resonances and since resonance strength falls with increasing order, I focus on resonances with $b$ or $q=2$.  To estimate the resonance locations I follow MP93 and use the mode frequencies from \citet{M91} for a standard Saturn interior model with no differential rotation and no composition gradients.  Because Marley did not consider oscillation modes that would not produce a first order resonance in the C- or D-rings, not all necessary mode frequencies were computed by him. To supplement the tabulated mode frequencies I measured f-mode frequencies from Figure 4 of \citet{VZ81} by digitizing the figure.

Table I tabulates the mode periods from \citet{M91} and \citet{VZ81} and the resonant locations, $R$, computed with Equations (1) and (2) assuming Ê $b$ and $q \le 2$. 
Since for an oblate planet $\Omega$ includes terms that vary with a power series in $R$, there is no analytic solution for resonance location, which instead must be found iteratively. Instead I simply truncate the expressions for $\Omega$ and $\kappa$ to include only the gravitational harmonic $J_2$ which gives radii accurate to a few percent. This is sufficient to predict the general location in the rings where the resonances lie. If the known ring waves are found to have the necessary characteristics to be produced by interactions with modes of Saturn then a more complex analysis would be required. The resonant locations are also shown in Figures 1 and 2 for density and vertical resonances, respectively. Given the age of the models and the approximate method used to compute the resonant locations, these locations should only be used as a guide to identify the regions of the C-ring where waves might be launched at second order resonances.

\section{Discussion}

Figures 1 and 2 illustrate the locations of the predicted first order resonance locations in the ring as delineated in Marley \& Porco (1993) (black points) and also the newly predicted locations for second order resonances (blue points). Only those modes with $\ell \le 8$ are shown since modes with higher $\ell$ were computed by neither Marley (1991) nor \citet{VZ81}. Thus while  first order ($q=1$) outer Lindblad resonances with $m=\ell-2$ and first order ($b=1$) outer vertical resonances with $m=\ell-3$ will fall within the inner C-ring for $\ell \sim 9$ to 10 the  locations cannot be computed without new oscillation mode calculations.  In the remainder of this section I focus on the second order resonances with $b=q=2$ which are found throughout the C-ring for $\ell \le 8$.

\subsection{Possible New Resonances}

Given the locations of the second order resonances a few trends stand out. Second order resonances in general do not lie near the first order resonances with the same value of $m$. 
Thus a second order resonance and a first order resonance--with the same $m$ but different $\ell$--seldom lie in close juxtaposition. There are two cases where second and first order resonances are in close proximity. The first order resonance for the f-mode with $\ell = m = 5$ lies very close to the second order resonance with $\ell =7$, $m=5$ (enclosed by a dashed box Figure 1). There is a similar case for $m=6$ (also boxed).  If two $m=5$ or 6 waves had been identified at these locations then the apparent splitting could be attributed to the overlap of two different resonances. However such close pairings of resonances with the same $m$ are rare. Certainly this is not the explanation for the multiple waves with the same value of $m$ observed for $m=2$ and $m=3$ by HN13.

More intriguing are the multitude of second order density and vertical resonances, particularly in the outer C-ring (Figures 1 and 2).  The $m=\ell-1$, $b=2$ outer vertical resonances fall near the ten outward propagating wave features identified by \citet{2011Icar..216..292B} that lie between about 85,000 and 90,000 km (Figure 2). Some or all of these waves, of course, could be density waves associated with external satellite resonances that have not yet been identified. But if any of these are bending waves, then they could not be launched by an external satellite and an association with the planetary modes would have to be considered.
If the mechanism that produces mode splitting also affects these particular modes then a handful of the modes might be responsible for several of the ring features.
Likewise second order OVR resonances for modes with $m=\ell-3$ fall in the inner C-ring and the D-ring.  

The second order resonances for those modes that apparently produce the six known C-ring density waves (HN13) lie in the B-ring (at 104,000 and 94,600 km) and the outermost C-ring. These regions should be studied for possible density wave features, although the greater surface mass density of the B-ring and the greater distance from Saturn may inhibit large amplitude responses.  Other locations of second order eccentric resonances (Figure 1) are the outer C-ring (for $m=\ell$) and the inner C-ring (for $m=\ell-2$). Some inward propagating, unassociated waves are found in both locations and these merit further study as well. The strongest second order C-ring OLR would be with the $\ell = 3$, $m=1$ f-mode which would fall near 76,000 km where several wave features are noted by \citet{2011Icar..216..292B}.

\subsection{Origin of Second Order Resonances}

For a second order resonance to be excited there must be some asymmetry in the perturbing external gravitational potential.  In the case of an external satellite, this arises from the eccentric and/or vertical motion of the perturbing moon.  In a simple example, at one point in its orbit a ring particle may align with an eccentric moon at apoapse. If the orbital frequency commensurability is such that the next alignment occurs with the moon at periapse, the two encounters are not exactly the same and there is thus a periodic resonant forcing in the rotating frame of the ring particle arising from the eccentric orbit of the moon. In the case of a planetary oscillation mode, the source of such an asymmetry is less obvious if it can arise at all. Without an asymmetry there cannot be a second order resonance. 

There are many potential sources of perturbation which conceivably might distort an oscillation mode in an analogous manner to an eccentric satellite. Possibilities might include diurnal or seasonal temperature variation in the atmosphere of Saturn, the constantly varying tidal bulges from the satellites, the magnetic field, and inhomogeneities in the internal density and thermal structure of the planet. A rigorous examination of the effect of such perturbations on the internal modes is well beyond the scope of this Note. However it is possible estimate the order of magnitude effect of several of these.

Saturn's atmospheric temperature profile is known to vary as a function of location, altitude, and time. Saturn's tropopause temperature varies between the northern and southern hemispheres, presumably as a consequence of seasonal radiative forcing. At $P\sim 0.1\,\rm bar$ there is a 10 K difference between $30^\circ$ N and S latitude while the stratosphere of Saturn has been observed to cool by a similar amount at certain latitudes just between 2004 and 2009 \citep{2010Icar..208..337F}. This means that the resonant cavity in which many Saturnian oscillation modes are trapped is not N-S symmetric. However oscillation modes with periods exceeding about 15 minutes are trapped below the 1-bar level on Saturn \citep{1994A&A...291.1019M} where the hemispheric temperature contrast is far less \citep{2010Icar..208..337F}. Thus, for the f-mode oscillations considered here, thermal perturbations at the tropopause likely play an insignificant role, a conclusion supported by detailed calculations by \citet{1994A&A...291.1019M} (see their Figure 4 for $n=0$). Because radiative time constants for the deep atmosphere are  very long, diurnal temperature perturbations certainly are insignificant and thus it  there is likely no E-W asymmetry at depth either. Thus shallow temperature perturbations are an unlikely perturber.

Thermal anisotropies deeper in the planet are also conceivable. \citet{2011exop.book..471S} give scaling laws for estimating the thermal and density perturbations arising from convection in the interiors of giant planets.  They estimate that fractional density perturbations arising from convection in the deep interior are quite small, of order $10^{-8}$. Anisotropies in the internal convection pattern, if present at all, are unlikely to substantially perturb global free oscillations.

The gravitational bulges on Saturn induced by the gravity of external satellites distort the interior structure in a non-axisymmetric pattern. Tidal distortion can be estimated from standard theory (e.g., Hubbard (1985), eq. 4-58). Amplitudes of the tidal perturbation are likely small, of order a meter or so for Titan and less for the icy satellites. While small, this tidal amplitude is comparable to the f-mode oscillation amplitude required to excite ring waves \citep{MP93} so it is plausible that the tidal bulges could play a role in deforming the density perturbation pattern within the planet sufficiently to allow for second order resonances to be permitted. More work is needed to further determine the plausibility of this mechanism.

One other possible source of asymmetry could arise from mode interactions. The mode eigenfunctions describing the periodic displacement in the interior of a rotating planet, 
${\bf u}_{\ell,m,n}$ must be computed as a series, accounting for interaction of different modes.  Mode mixing and non-linear coupling is discussed in some detail by \citet{Ful13} but here we use the nomenclature and solution of \citet{VZ81} who only considered mixing arising from Coriolis and ellipticity effects. They find that
$${\bf u}_{\ell,m,n} = \sum_{k=0}^2 c_k{\bf u}_{\ell_k, m, n_k}^0\eqno(4)$$ 
where a mode with $(\ell, m, n)$ can be described as the sum of three interacting modes, all with the same value of $m$ but with three different pairs of $\ell$ and $n$, one of which is the unperturbed mode $(\ell_0,m,n_0)$. The weighting function $c_k$ gives the projection of each of the three interacting modes computed for a non-rotating planet, ${\bf u}^0_{\ell,m,n}$ on the observed mode.

If the relative contribution of each of the three interacting modes varies with time then $c_k = c_k(t)$, and for a fixed $\ell, m, n$ the internal total density perturbation arising from ${\bf u}_{\ell,m,n}$
along with its associated perturbation to the external gravitational potential will vary with time in the co-rotating frame. Time variability in the mode mixing might be induced by perturbations such as those discussed above or by influences not yet recognized.
If present this would be a source of asymmetric perturbations on Saturn's external gravitational field similar to those associated with an eccentric moon, and would ikewise  allow for second order resonances in the rings. 

The discussion in this section is simply meant to argue that there could be oscillation mode anisotropies that would permit second order resonances to form. 
Unlike the highly idealized case of most oscillating stars, an oblate Saturn with a core, hydrogen phase transition, helium rain, and multiple tidal bulges could well be an imperfect resonator. Whether or not these imperfections can indeed lead to additional resonance features can best the ascertained by careful study of the currently known C-ring wave features. If the properties of the observed wave features are consistent with forcing at outer Lindblad or outer vertical second order resonances, the mechanism will require much more detailed investigation.

\section{Conclusion}
Analysis of Cassini stellar occultation data supports the hypothesis of Marley (1990) and Marley and Porco (1993) that periodic perturbations of the external gravitational field of Saturn induced by planetary oscillation modes resonantly perturb ring particle orbits, thereby launching waves in the C-ring of Saturn. 
Given the novelty of the proposed mechanism those authors focused entirely on first order resonances between planetary modes and ring particle orbits. However second order resonances between low $\ell$ f-mode oscillations of Saturn and ring orbits also lie in the C-ring in locations where waves are found by \citet{2011Icar..216..292B}  which are not associated with known satellite resonances. At least two resonant locations fall in the inner regions of the B-ring. Further analysis of the known waves similar to that carried out by \citet{HN13}  would ascertain if these features are also associated with oscillation modes. If so, then these resonances would provide measurements of additional planetary oscillation frequencies for modes not currently measured and thus new constraints on the interior structure of Saturn. Since the current associations only constrain three modes, each additional wave feature provides a valuable new constraint on Saturn's interior structure.  

Second order resonances are not, however, responsible for the observed ``fine'' mode splitting discovered by \citet{HN13}.  The mechanism behind this splitting has thus not yet been identified.

\section{Acknowledgements} I thank Jeff Cuzzi and Jack Lissauer for helpful discussions and comments on this manuscript. I also gratefully acknowledge the reviewers for thoughtful and helpful comments. This work was supported by the NASA OPR Program.

\newpage

\begin{table}

\centerline{Table I}

    \begin{tabular}{cccccccl}
     \hline
       \multicolumn{3}{c}{mode}  & \multicolumn{4}{c}{Resonance Locations$^\dag$}  \\
 $\ell$ & $m$  & $P_{\rm pat}^\ddag$  & OLR ($q=1$)  & OLR ($q=2$)  & OVR ($b=1$)   & OVR ($b=2$)       \\  
   ~ & ~   &  (min)  &  (km) &   (km) & (km) & (km)    \\ \hline

    2 & 2 & 286.1   & 86,140         & 104,180         & ~         & ~             \\
    ~ & 1 & 150.1   & ~         & ~         &69,170        & 91,110            \\ \hline
    3 & 3 & 296.7   & 81,640        & 94,600        & ~         & ~            \\
    ~ & 2 & 211.6   & ~         & ~         & 71,410         & 86,950              \\
    ~ & 1 & 118.2(v)   & --         & 75,590        & ~         & ~             \\ \hline
    4 & 4 & 310.7    & 80,700         & 91,010       & ~         & ~            \\
    ~ & 3 & 249.3    & ~         & ~         &73,460         & 85,650             \\
    ~ & 2 & 183.9(v) & 64,150        & 77,590         & ~         & ~            \\ 
    ~ & 1 & 100.1(v)    & ~         & ~         &--         & 69,530             \\ \hline
    5 & 5 &324.2   & 80,810        & 89,450         & ~         & ~             \\
    ~ & 4 & 276.3    & ~         & ~         & 75,240         & 85,320             \\
    ~ & 3 & 223.3    & 67,550         & 78,280         & ~         & ~             \\ 
    ~ & 2 & 163.9(v)    &          & ~         & --        & 73,350           \\ \hline
    6 & 6 & 335.6   & 81,180         & 88,640         & ~         & ~             \\
    ~ & 5 & 297.5   & ~         & ~         &76,830        & 85,470             \\
    ~ & 4 & 254.2    & 70,580         & 79,600         & ~         & ~            \\  
    ~ & 3 & 206.8(v)   &           & ~         & 64,852         & 75,610            \\  
    ~ & 2 & 150.0(v)    & --         & 67,730         & ~         & ~            \\ \hline
    7 & 7 & 348.6(v)    & 82,130        & 85,810         & ~         & ~           \\
    ~ & 6 & 313.5    & ~         & ~         & 78,030         & 85,580             \\  
    ~ & 5 & 277.8    & 72,890        & 80,700         & ~         & ~             \\  
    ~ & 4 & 238.0    & ~         & ~         & 68,110         & 77,240             \\  
    ~ & 3 & 192.9(v)    & --         & 71,000         & ~         & ~             \\ \hline
    8 & 8 & 355.0    & 82,270         & 88,180         & ~         & ~           \\
    ~ & 7 & 327.3    & ~         & ~         & 79,150         & 85,880             \\     
    ~ & 6 & 297.5    & 74,900         & 81,800         & ~         & ~             \\ 
    ~ & 5 & 263.8    & ~         & ~         & 70,910         & 78,880             \\ 
    ~ & 4 & 225.7    & 65,200         & 73,540         & ~         & ~             \\ \hline

    \end{tabular}
    \caption{$^\dag$Only resonances with $R> 64,000\,\rm km$ are shown. $^\ddag$Pattern periods from Marley (1991) except those denoted (v) which are from \citet{VZ81}.}
\end{table}

\newpage
\begin{figure} 
   \centering
\includegraphics[width=1.0\textwidth]{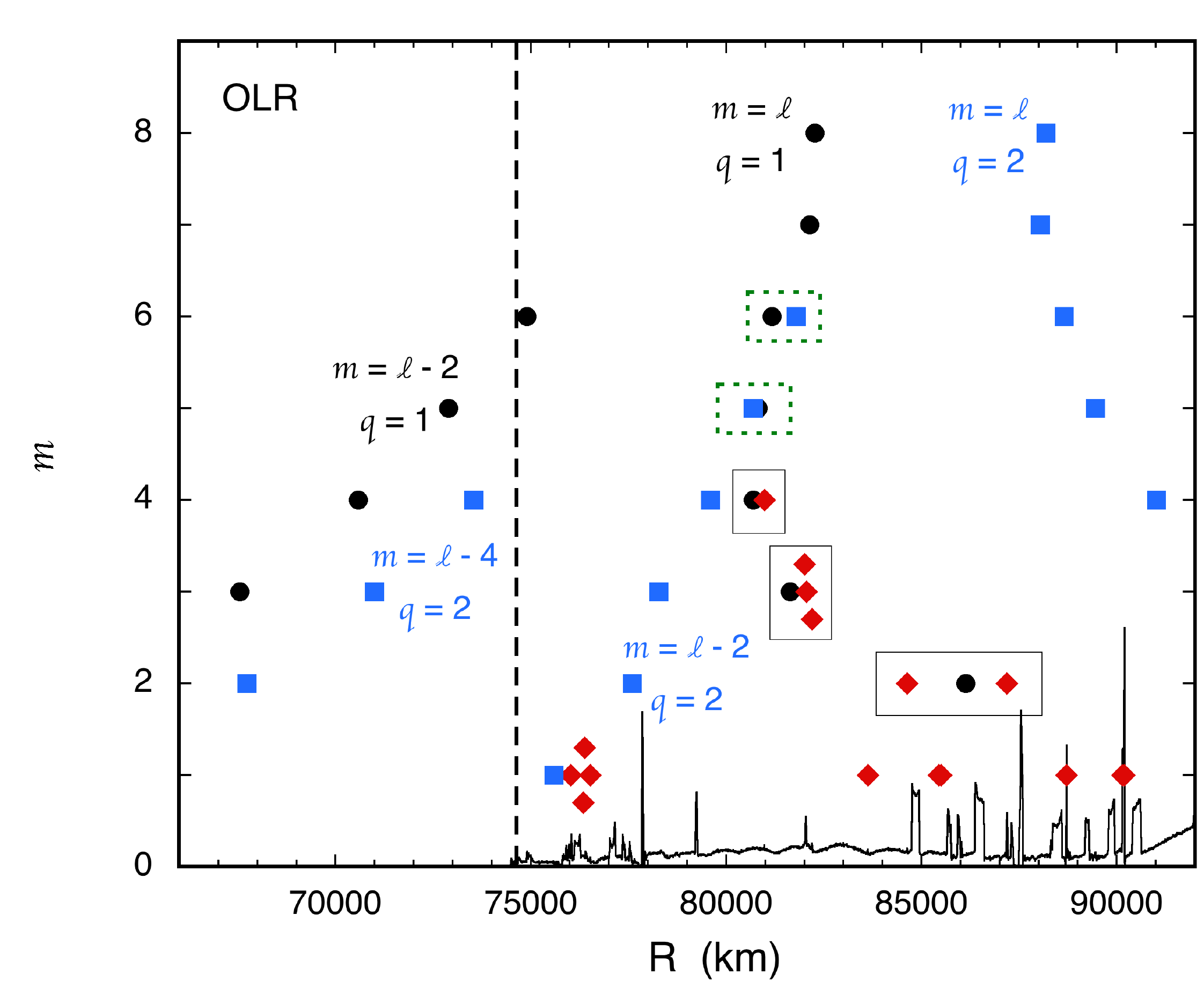} 
   \caption{Symbols denote the location and azimuthal wavenumber $m$ of various wave features in the C and D-rings. Observed ring wave features (HN13 and \citet{2011Icar..216..292B}) are shown by the red diamonds, slightly offset for clarity where necessary. Only waves that propagate towards Saturn are shown. Waves for which $m$ has not been measured are arbitrarily plotted at $m=0.5$. Predicted f-mode resonance locations for first (second) order outer resonances are shown by black circles (blue squares). For the first order case ($q=1$) those resonances with  modes having $m=\ell$ and $m=\ell-2$ are shown. For the second order resonances ($q=2$) the same two cases are shown as well as for $m=\ell-4$. The vertical line denotes the boundary between the D- and C-rings. Solid boxes group waves and mode resonances that appear \citep{HN13} to be associated. Dashed boxes highlight the only two cases where first and second order resonance locations nearly overlap. A map of C-ring UV opacity as a function of radius (courtesy J.\ Colwell, personal communication) is shown along bottom with an arbitrary linear vertical scale. }
\label{OLR}
\end{figure}
\newpage
\begin{figure} 
   \centering
\includegraphics[width=1.0\textwidth]{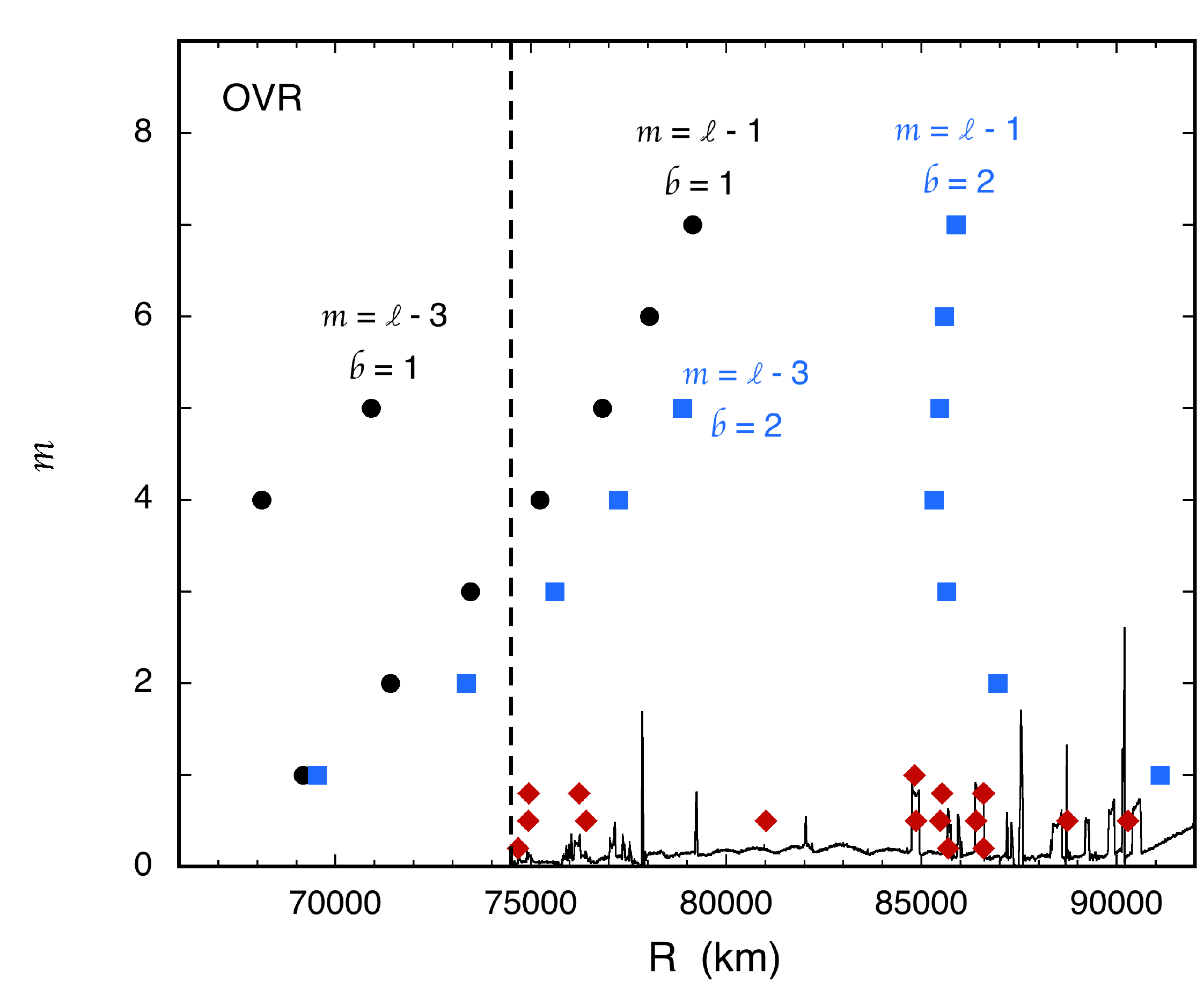} 
   \caption{Similar to Figure 1, the symbols denote the location of various observed and predicted ring wave features. Here only waves that propagate away from Saturn \citep{2011Icar..216..292B} are shown. There is no measured value of $m$ for any of these waves and they are arbitrarily plotted at $m=0.5$. First and second order resonances for two sets of f-modes ($m=\ell-1$ and $m=\ell-3$) are denoted by black circles and blue squares, respectively. Other plot details as in Figure 1. }
\label{OVR}
\end{figure}

\end{document}